\documentclass{llncs}
\usepackage{color}
\usepackage{times}
\usepackage{graphicx}
\usepackage{latexsym}

\usepackage{caption}
\usepackage{amssymb,amsmath}
\usepackage{subfigure}
\usepackage{chngcntr}
\usepackage[linesnumbered]{algorithm2e}
\usepackage{multicol}
\usepackage{tikz,hyperref}

\definecolor{darkblue}{rgb}{0.1,0.1,0.4}

\newcommand \MI[3]{MI^{#3}(#1,#2) }
\newcommand \SL[2]{Slack(#1,#2) }

\newcommand \SOI[1]{SoC_{#1} }
\newcommand \DOI[1]{EoC_{#1}}

\begin{document}
\bibliographystyle{plain}

\title{The Energetic Reasoning Checker Revisited}

\author{Alban Derrien \and Thierry Petit\institute{
TASC (Mines Nantes, LINA, CNRS, INRIA), \\
4, Rue Alfred Kastler,
FR-44307~Nantes Cedex 3, France. \\
\{alban.derrien, thierry.petit\}@mines-nantes.fr 
}
}

\maketitle
\sloppy 

\begin{abstract} 
Energetic Reasoning  (ER) is a powerful filtering algorithm for the Cumulative constraint. Unfortunately, ER is generally too costly to be used in practice. 
One reason of its bad behavior is that many intervals are considered as relevant by the checker of ER, although most of them should be ignored. In this paper, we provide a sharp characterization that 
allows to reduce the number of intervals by a factor seven. Our experiments show that associating this checker with a Time-Table 
filtering algorithm leads to promising results. 
\end{abstract}

\section{Introduction}
Due to its relevance in many industrial contexts, the Cumulative Scheduling Problem (CuSP) has been widely studied in Constraint Programming (CP). 
The CuSP is NP-Hard. It is defined by a set of activities $\mathcal{A}$ and a resource of capacity $C$. 
Each activity $a \in \mathcal{A}$ is defined by four variables: its starting time $s_a$, its processing time $p_a$, its ending time $e_a$ and its height $h_a$, which represents the amount of resources consumed by the activity 
when it is processed. We use the notation $a = \{ s_a, p_a, e_a, h_a  \}$. 
Usually, variables $p_a$ and $h_a$ are fixed integers. In this paper, we make such an assumption. A solution to a CuSP is a schedule that satisfies the following constraints: 
\begin{center}
$\forall a \in \mathcal{A}: s_a + p_a = e_a \notag$~~~~~~$\wedge$~~~~~~
$\forall i \in \mathbb{N}: \sum \limits_{  \underset{i \in [s_a, e_a[}{a \in \mathcal{A}} } h_a \leq C \notag$ 
\end{center}

In CP, this problem is generally represented by the global constraint \textsf{Cumulative} ~\cite{aggbel93}.
The Energetic Reasoning of Baptiste et al. (ER) is one of the most powerful filtering algorithm for Cumulative~\cite{baptiste:inria-00123562}. 
This algorithm uses a characterization of intervals of interest, that is, intervals that are sufficient to check in order to ensure that all the undergoing rules used 
for filtering domains are satisfied. Unfortunately, ER is often too costly to be used in practice. First, its time complexity is $O(n^3)$. Moreover, 
the hidden constant in that time complexity is huge, as many intervals are characterized to be of interest although most of them should be ignored. 
In the literature, only heuristics approaches have been proposed for reducing the number of checked intervals~\cite{Berthold:2011:ApproxCrit}. 

This article provides a sharper characterization of intervals of interest. 
Our experiments show a significant reduction in the running time of the ER checker. We point out that assocating this energetic checker with a Time-Table filtering algorithm (based on the cumulative profile of mandatory parts~\cite{Letort:ScalableSweep})
leads to promising results. Further, our approach should permit to answer affirmatively to a theoretical open question with respect to energetic filtering algorithms: 
Are the intervals of interest for the ER checker sufficient in order to perform a complete ER filtering for the activities variables?
\section{Background}\label{sec:background}
Given a variable $x$,
we denote $\underline{x}$ the minimum value in its domain and $\overline{x}$ the maximum value. The principle of ER
is to compare the available area within a given interval (length of that interval $\times$ capacity) with the quantity of resource necessarily taken by 
activities that should partially or totally overlap this interval. The ER checker is defined using $\MI{t_1}{t_2}{a}$, the \emph{minimum intersection} of activity $a$ with an interval $[t_1,t_2]$. 
$\MI{t_1}{t_2}{a} = \max\left(0,\min\left(p_a,~t_2\!-\!t_1,~\underline{e_a} \!-\! t_1,~t_2\!-\!\overline{s_a}\right)\right)$.

 \begin{proposition}[ER checker~\cite{ers89:ER}]\label{prop:ER}
 If the condition 
 \begin{eqnarray}
 \forall t_1,t_2 \in \mathbb{N}^2,t_1<t_2 \hspace{.5cm} C\times(t_2-t_1) \geq \sum \limits_{a \in \mathcal{A}} h_a \times   \MI{t_1}{t_2}{a}\label{cond:prop:ER}
 \end{eqnarray} 
 is violated then the problem represented by \emph{Cumulative} is unfeasible.
 \end{proposition}
The issue is then to find the smallest set of intervals $[t_1,t_2], t_2 > t_1$ that should be checked to detect the unfeasibility.
From this condition a bound adjustment rule can be defined.
Value $\underline{s_a}$ can be updated if scheduling activity $a$ on $\underline{s_a}$ would cause an overload as described above.
To check such potential overloads, we need to compute how much additional energy $a$  would require during $[t_1,t_2]$ if $a$ is scheduled on $\underline{s_a}$.

\paragraph{\bf Baptiste et al. characterization}  
In order to ensure that condition (\ref{cond:prop:ER}) holds, it is enough to check intervals of the form: 
    $[t_1,t_2], t_1 \in O_1 < t_2 \in O_2$,~~~
    $[t_1,t_2], t_1 \in O_1 < t_2 \in O(t_1)$,~~~
    $[t_1,t_2], t_2 \in O_2 > t_1 \in O(t_2)$,
where
$O_1 = \{\underline{s_a}, \forall a \in\mathcal{A}\}    \cup   \{\overline{s_a}, \forall a \in\mathcal{A}\}   \cup   \{\underline{e_a}, \forall a \in\mathcal{A}\} $, 
$O_2 = \{\overline{e_a}, \forall a \in\mathcal{A}\}    \cup   \{\overline{s_a}, \forall a \in\mathcal{A}\}   \cup   \{\underline{e_a}, \forall a \in\mathcal{A}\}  $,
$O(t)  = \{\underline{s_a} + \overline{e_a} - t, \forall a \in\mathcal{A}\} $.
The resulting subset of intervals forms the \emph{intervals of interest} of the ER. 
There is for each pair of activities $9\!+\!3\!+\!3\!=\!15$  intervals of interest. 
The characterization is proved to be sufficient in~\cite{baptiste:inria-00123562} (proposition 19), by implicitly analyzing the slack function
 $\SL{t_1}{t_2} = C\times(t_2-t_1) - \sum_{a \in \mathcal{A}} h_a \times   \MI{t_1}{t_2}{a}$.
As $\MI{t_1}{t_2}{a}$ can be computed in constant time and given the definitions of $O_1$, $O_2$ and $O_t$, we obtain a naive checker in $O(n^3)$, by computing $\MI{t_1}{t_2}{a}$ for all $a,t_1,t_2$. 
However, the slack function is continuous piecewise linear, a local extrema can only be found at flexion points,  i.e., when the slope is changing, which has been proved to be only in such intervals. 
This leads to a checker in $O(n^2)$~\cite{baptiste:inria-00123562}.
Two open questions remain.
\begin{enumerate}
\item The new set of intervals of interest is proved to be sufficient but could it be reduced?
\item Is the new set of intervals also sufficient to ensure a complete filtering of the time bound adjustment rule based on the checker? 
\end{enumerate}

\paragraph{\bf Schwindt's characterization} Schwindt has proposed a finer characterization of the minimal value of the slack function (Theorems 3.7 and 3.8 in~\cite{schwindt98:PhD}, written in German):
The slack function is locally minimal in interval $[t_1,t_2]$ only if its left derivative is negative and  right derivative positive, for both $t_1$ and $t_2$.
As the slack function correspond to the negation of the sum of minimal intersections,
it must exists an activity $i$ (resp $j$) such that its minimal intersection has a left derivative greater than its right derivative on $t_1$ (resp. $t_2$). 
This leads to Theorem \ref{theo:existsLandRderivate}.

\begin{theorem}\label{theo:existsLandRderivate}
The slack function is locally minimum in interval $[t_1,t_2]$ only if it exists two activities $i,j$ such that the two following conditions are satisfied. 

\begin{align} 
 \frac{\delta^- \MI{t_1}{t_2}{i}}{\delta t_1} > \frac{\delta^+ \MI{t_1}{t_2}{i}}{\delta t_1}  \label{cond:derivativeLT1}\\
\frac{\delta^- \MI{t_1}{t_2}{j}}{\delta t_2} > \frac{\delta^+ \MI{t_1}{t_2}{j}}{\delta t_2}  \label{cond:derivativeLT2}
\end{align} 
\end{theorem}

Schwindt analyzes the variation of minimum intersection. He provides a first characterization of the 8 possible intervals of interest for any pair of activities. 
This answers to the first open question: The number of intervals in Baptiste et al. characterization can be reduced.
We propose in next section a different analyze of the minimum intersection, which leads to a sharper characterization of activity pairs.

\section{A New Characterization}

For symmetry reasons we focus on condition (\ref{cond:derivativeLT2}).
We wish to characterize for a starting time $t_1$ and an activity $a$ the positive inflection points of function $t_2 \rightarrow \MI{t_1}{t_2}{a}$:
Values $t_2$ at which the left derivative of is greater than the right derivative.


\begin{theorem}\label{theo:maxOneInflection}
For a starting time $t_1$ and an activity $a$ the function  $t_2 \rightarrow \MI{t_1}{t_2}{a}$ has at most one positive inflection point.
\end{theorem}
{\small
\begin{proof}
We  prove theorem \ref{theo:maxOneInflection} by studying the four positions of $t_1$ w.r.t. $a$.
We prove that the function $t_2 \rightarrow \MI{t_1}{t_2}{a}$ is continuous piecewise linear, composed of at most three parts.
The two inflection points correspond to the start of consumption ($\SOI{a}$) 
and the end of consumption ($\DOI{a}$). We also prove that $\DOI{a}$ is the only inflection point with a left derivative greater than the right derivative.
Graphically we show an example of each cases with an activity $a = \{ s_a \!\in\! [2,4],~ p_a\!=\!4,~e_a\!\in\![6,8],h_a\}$.

\paragraph{Case 1.}
\paragraph{}
\hspace{-0.15cm}\begin{minipage}{0.45\textwidth} 
On the first case, the minimum starting time of the activity $a$ is greater than or equal to the profile starting time: 
$t_1 \le \underline{s_a} $.
\end{minipage}
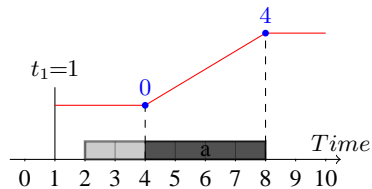
\begin{minipage}{0.45\textwidth} 
   \hspace{0.5cm}
\begin{tikzpicture}[xscale=0.4,yscale=0.24]
  \draw[->] (-0.5,0) -- ( 10.5 , 0 ) node[above] {$Time$};    
\foreach \x in {0,...,10}\draw (\x,1pt) -- (\x,-3pt) node[anchor=north] {\x};

  \draw (1,0) -- ( 1 , 4 ) node[above] {$t_1\!\!=\!\!1$};  
\pgfmathsetmacro{\CoinBasX}{2}\pgfmathsetmacro{\CoinBasY}{0}
\pgfmathsetmacro{\CoinHautX}{8}\pgfmathsetmacro{\CoinHautY}{1}
\pgfmathsetmacro{\Duration}{4}
\pgfmathsetmacro{\Delta}{1}
\foreach \x in {\CoinBasX,...,\CoinHautX}
	\draw[gray] (\x,\CoinBasY) -- (\x,\CoinHautY);
 \filldraw[thin,fill=gray,opacity=.4,draw=black,line width=1pt] (\CoinBasX, \CoinBasY)   rectangle (\CoinHautX,\CoinHautY);
 \filldraw[thin,black,opacity=.6] (\CoinBasX/2+\CoinHautX/2 +\Delta-\Duration/2, \CoinBasY)   rectangle (\CoinBasX/2+\CoinHautX/2 +\Delta + \Duration/2,\CoinHautY);
 \node[black]  at ( \CoinBasX/2+\CoinHautX/2 +\Delta , \CoinBasY/2 +\CoinHautY /2) {a } ;
 
 \pgfmathtruncatemacro{\Delt}{3}  
  \draw[red] (1,\Delt+0) --(2,\Delt+0) --(3,\Delt+0) --(4,\Delt+0) --(5,\Delt+1) --(6,\Delt+2) -- (7,\Delt+3) -- (8,\Delt+4) --(9,\Delt+4) --(10,\Delt+4) 	; 
\draw[dashed]	   (4,0) -- (4,\Delt+0) node[fill=blue,circle,scale=0.3] {};
\draw[dashed]	   (8,0) -- (8,\Delt+4) node[fill=blue,circle,scale=0.3] {};
\node[blue]  at ( 4, \Delt +1 ) {$0$} ;
\node[blue]  at ( 8, \Delt  +5) {$4$} ;
 \end{tikzpicture} 
\label{Fig:case1}
\end{minipage}\\
By definition we have $ \MI{t_1}{t_2}{a} = \max\left(0,\min\left(p_a,~t_2\!-\!t_1,~\underline{e_a} \!-\! t_1,~t_2\!-\!\overline{s_a}\right)\right)$.
We can deduce three different situations:

\begin{enumerate}
\item if $t_2 \le \overline{s_a}$ then $ \MI{t_1}{t_2}{a}= 0$.
\item or $\overline{s_a}\le t_2 \le \overline{e_a}$ then $\MI{t_1}{t_2}{a}=  t_2 - \overline{s_a}$.
\item and finally $ \overline{e_a} \le t_2$ then $\MI{t_1}{t_2}{a} = p_a$.
\end{enumerate}

\noindent $p_a \!-\! (\overline{e_a} \!-\! t_2)$ equals $0$ when $t_2 = \overline{s_a}$ and $p_a$ when $t_2 = \overline{e_a}$.
This proves that when $t_1 \le \underline{s_a} $ then consumption function is continue and piecewise linear, composed of three pieces with only one positive inflection point: $\DOI{a}$. 

\paragraph{Case 2.}
\paragraph{}
\hspace{-0.15cm}\begin{minipage}{0.45\textwidth} 
On the second case, the minimum ending time of the activity $a$ is smaller than or equal to the profile starting time:
$t_1 \ge \underline{e_a} $.
\end{minipage}
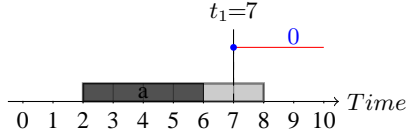
\begin{minipage}{0.45\textwidth} 
   \hspace{0.5cm}
\begin{tikzpicture}[xscale=0.4,yscale=0.24]
  \draw[->] (-0.5,0) -- ( 10.5 , 0 ) node[right] {$Time$};    
\foreach \x in {0,...,10}\draw (\x,1pt) -- (\x,-3pt) node[anchor=north] {\x};

  \draw (7,0) -- ( 7, 4 ) node[above] {$t_1\!\!=\!\!7$};  
  \pgfmathsetmacro{\CoinBasX}{2}\pgfmathsetmacro{\CoinBasY}{0}
\pgfmathsetmacro{\CoinHautX}{8}\pgfmathsetmacro{\CoinHautY}{1}
\pgfmathsetmacro{\Duration}{4}
\pgfmathsetmacro{\Delta}{-1}
\foreach \x in {\CoinBasX,...,\CoinHautX}
	\draw[gray] (\x,\CoinBasY) -- (\x,\CoinHautY);
 \filldraw[thin,fill=gray,opacity=.4,draw=black,line width=1pt] (\CoinBasX, \CoinBasY)   rectangle (\CoinHautX,\CoinHautY);
 \filldraw[thin,black,opacity=.6] (\CoinBasX/2+\CoinHautX/2 +\Delta-\Duration/2, \CoinBasY)   rectangle (\CoinBasX/2+\CoinHautX/2 +\Delta + \Duration/2,\CoinHautY);
 \node[black]  at ( \CoinBasX/2+\CoinHautX/2 +\Delta , \CoinBasY/2 +\CoinHautY /2) {a } ;

 \pgfmathtruncatemacro{\Delt}{3}  
  \draw[red] (7,\Delt+0) -- (8,\Delt+0) --(9,\Delt+0) --(10,\Delt+0) 	; 
\draw[dashed]	   (7,0) -- (7,\Delt+0) node[fill=blue,circle,scale=0.3] {};
\node[blue]  at ( 9, \Delt +0.6 ) {$0$} ;
 \end{tikzpicture} 
\label{Fig:case2}
\end{minipage}\\
\noindent In this case, we have $\MI{t_1}{t_2}{a} = 0$ for any interval. Then the function is trivially continue and piecewise linear with zero inflection point. 
\paragraph{Case 3.}
\paragraph{}
\hspace{-0.15cm}\begin{minipage}{0.45\textwidth} 
On the third case, the profile starting time is greater than the minimum start time, but smaller than the minimum end time and maximum start time:\\
  $ t_1 \!>\! \underline{s_a} 	~~and~~  t_1 \!<\!  \underline{e_a}	~~and~~ 	t_1 \! <\!  \overline{s_a} $
\end{minipage}
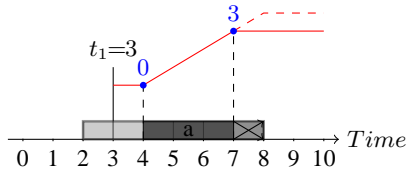
\begin{minipage}{0.45\textwidth} 
   \hspace{0.5cm}
\begin{tikzpicture}[xscale=0.4,yscale=0.24]
  \draw[->] (-0.5,0) -- ( 10.5 , 0 ) node[right] {$Time$};
    
\foreach \x in {0,...,10}\draw (\x,1pt) -- (\x,-3pt) node[anchor=north] {\x};

  \draw (3,0) -- ( 3 , 4 ) node[above] {$t_1\!\!=\!\!3$};  
\pgfmathsetmacro{\CoinBasX}{2}\pgfmathsetmacro{\CoinBasY}{0}
\pgfmathsetmacro{\CoinHautX}{8}\pgfmathsetmacro{\CoinHautY}{1}
\pgfmathsetmacro{\Duration}{3}
\pgfmathsetmacro{\Delta}{.5}
\foreach \x in {\CoinBasX,...,\CoinHautX}
	\draw[gray] (\x,\CoinBasY) -- (\x,\CoinHautY);
 \filldraw[thin,fill=gray,opacity=.4,draw=black,line width=1pt] (\CoinBasX, \CoinBasY)   rectangle (\CoinHautX,\CoinHautY);
 \filldraw[thin,black,opacity=.6] (\CoinBasX/2+\CoinHautX/2 +\Delta-\Duration/2, \CoinBasY)   rectangle (\CoinBasX/2+\CoinHautX/2 +\Delta + \Duration/2,\CoinHautY);
 \node[black]  at ( \CoinBasX/2+\CoinHautX/2 +\Delta , \CoinBasY/2 +\CoinHautY /2) {a } ;
 
 \filldraw[thin,black,opacity=.3] (7,0)   rectangle (8,1);
  \draw[->] (7,0) -- ( 8 , 1 ) ;
  \draw[->] (7,1) -- ( 8 , 0 ) ; 
 \pgfmathtruncatemacro{\Delt}{3}  
  \draw[red] (3,\Delt+0) --(4,\Delt+0) --(5,\Delt+1) --(6,\Delt+2) -- (7,\Delt+3) --(10,\Delt+3) 	; 
  \draw[red,dashed]  (7,\Delt+3) -- (8,\Delt+4) --(10,\Delt+4) 	; 
\draw[dashed]	   (4,0) -- (4,\Delt+0) node[fill=blue,circle,scale=0.3] {};
\draw[dashed]	   (7,0) -- (7,\Delt+3) node[fill=blue,circle,scale=0.3] {};
\node[blue]  at ( 4, \Delt +1 ) {$0$} ;
\node[blue]  at ( 7, \Delt  +4) {$3$} ;
 \end{tikzpicture} 
\label{Fig:case3}
\end{minipage}\\

\noindent Let $\Delta = t_1-\underline{s_a} $. We distinct three cases for the value of $t_2$ :
\begin{enumerate}
\item if $t_2 \le \overline{s_a}$ then $\MI{t_1}{t_2}{a}=0$.
\item or $\overline{s_j} \le t_2 \le \overline{e_j} -\Delta$ then $\MI{t_1}{t_2}{a}= t_2- \overline{s_a}$.
\item and finally $t_2 \ge \overline{e_a} -\Delta$ then$\MI{t_1}{t_2}{a} = p_a-\Delta$.
\end{enumerate}
As $t_2- \overline{s_a}$ equals $0$ when $t_2 = \overline{s_a}$ and $p_a-\Delta$ when $t_2 = \overline{e_a} -\Delta$ we have proved that the consumption is continue and piecewise linear, composed of three pieces with only one positive inflection point: $\DOI{a}$.

\paragraph{Case 4.}
\paragraph{}
\hspace{-0.15cm}\begin{minipage}{0.45\textwidth} 
On the fourth case, the profile starting time intersect the mandatory part $a$:\\
  $ t_1 \!>\! \underline{s_a} 	~~and~~  t_1 \!<\!  \underline{e_a}	~~and~~ 	t_1 \! \ge\!  \overline{s_a} $ 
\end{minipage}
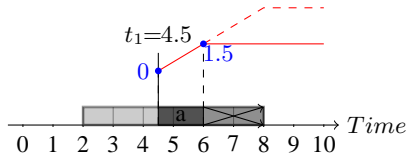
\begin{minipage}{0.45\textwidth} 
   \hspace{0.5cm}
\begin{tikzpicture}[xscale=0.4,yscale=0.24]
  \draw[->] (-0.5,0) -- ( 10.5 , 0 ) node[right] {$Time$};
    
\foreach \x in {0,...,10}\draw (\x,1pt) -- (\x,-3pt) node[anchor=north] {\x};

  \draw (4.5,0) -- ( 4.5 , 4 ) node[above] {$t_1\!\!=\!\!4.5$};
\pgfmathsetmacro{\CoinBasX}{2}\pgfmathsetmacro{\CoinBasY}{0}
\pgfmathsetmacro{\CoinHautX}{8}\pgfmathsetmacro{\CoinHautY}{1}
\pgfmathsetmacro{\Duration}{1.5}
\pgfmathsetmacro{\Delta}{0.25}
\foreach \x in {\CoinBasX,...,\CoinHautX}
	\draw[gray] (\x,\CoinBasY) -- (\x,\CoinHautY);
 \filldraw[thin,fill=gray,opacity=.4,draw=black,line width=1pt] (\CoinBasX, \CoinBasY)   rectangle (\CoinHautX,\CoinHautY);
 \filldraw[thin,black,opacity=.6] (\CoinBasX/2+\CoinHautX/2 +\Delta-\Duration/2, \CoinBasY)   rectangle (\CoinBasX/2+\CoinHautX/2 +\Delta + \Duration/2,\CoinHautY);
 \node[black]  at ( \CoinBasX/2+\CoinHautX/2 +\Delta , \CoinBasY/2 +\CoinHautY /2) {a } ;
 
 \filldraw[thin,black,opacity=.3] (6,0)   rectangle (8,1);
  \draw[->] (6,0) -- ( 8 , 1 ) ;
  \draw[->] (6,1) -- ( 8 , 0 ) ;

 \pgfmathtruncatemacro{\Delt}{3}
  \draw[red] (4.5,\Delt) --(5,\Delt+0.5) --(6,\Delt+1.5) -- (7,\Delt+1.5) --(10,\Delt+1.5) 	; 
  \draw[red,dashed]  (6,\Delt+1.5) -- (7,\Delt+2.5) -- (8,\Delt+3.5) --(10,\Delt+3.5) 	; 
\draw[dashed]	   (4.5,0) -- (4.5,\Delt) node[fill=blue,circle,scale=0.3] {};
\draw[dashed]	   (6,0) -- (6,\Delt+1.5) node[fill=blue,circle,scale=0.3] {};
\node[blue]  at ( 4, \Delt ) {$0$} ;
\node[blue]  at ( 6.5, \Delt  +1) {$1.5$} ;
 \end{tikzpicture} 
\label{Fig:case4}
\end{minipage}\\
\noindent In this case we have two distinct  cases for the value of $t_2$ :
\begin{enumerate}
\item if $t_2 \le \underline{e_a}$  then $\MI{t_1}{t_2}{a}=t_2-t_1$
\item otherwise $\MI{t_1}{t_2}{a} = p_a-\Delta$
\end{enumerate}
As $t_2-t_1 = p_a-\Delta$ when $t_2 = \underline{e_a}$ then the consumption function is continue and piecewise linear,
composed of two pieces, with only one positive inflection point : $\DOI{a}$. 
\paragraph{Conclusion}
We have shown that for any starting value $t_1$, and any activity $a$  the function $t_2 \rightarrow \MI{t_1}{t_2}{a}$ is linear and piecewise continue,
with, in each case, at most one point at which the left  derivative is greater than the right derivate: $\DOI{a}$. This proves the Theorem.  
\qed
\end{proof}
}

\newcommand\myLe{\!\le\!}
\newcommand\myGe{\!\ge\!}
\newcommand\myWedge{~\wedge~}
\newcommand\myPlus{\!+\!}
\newcommand\myMoins{\!-\!}
{
\begin{table}[!ht]
\vspace{-1.0cm}
\begin{minipage}{\textwidth}
\centering
\captionof{table}{\label{tab:characterization}Intervals of interest for a pair of activities ($i$,$j$)}
\begin{tabular}{|lll|r|}\cline{1-3}
\linespread{0.8}
\scriptsize
conditions &~~~~& interval \\ \hline
\scriptsize
$\underline{s_i} \myLe \underline{s_j} \myWedge  \overline{e_j} \myGe \overline{e_i}$
&&   $[\underline{s_i},~ \overline{e_j}]$ & ~A\\
\scriptsize
$\underline{s_i} \myGe \underline{s_j} \myWedge 
  \underline{s_i} \myLe \underline{e_j} \myWedge 
  \underline{s_i} \myLe \overline{s_j}  \myWedge
  \underline{s_j} \myPlus \overline{e_j} \myMoins \underline{s_i} \myGe \overline{e_i}$
&&   $[ \underline{s_i}   ,~   \underline{s_j} \myPlus \overline{e_j} \myMoins \underline{s_i}]$& B\\
\scriptsize
$\underline{s_i} \myGe \underline{s_j} \myWedge 
  \underline{s_i} \myLe \underline{e_j} \myWedge 
  \underline{e_j} \myGe  \overline{e_i}$
&& $[ \underline{s_i}   ,~   \underline{e_j}]$& C\\
\scriptsize
$\overline{s_i} \myLe \underline{s_j} \myLe \overline{e_j} \myLe \underline{e_i}  $
&&$[ \overline{s_i}   ,  \overline{e_j}]$ &D\\
  \scriptsize
$\overline{s_i} \myGe \underline{s_j} \myWedge 
\overline{s_i} \myLe \underline{e_j} \myWedge 
\overline{s_i} \myLe \overline{s_j}  \myWedge
\underline{s_j} \myPlus \overline{e_j} \myLe \overline{s_i} \myPlus \underline{e_i} \myWedge
\underline{s_j} \myPlus \overline{e_j} \myGe 2 \times \overline{s_i}$
&& $[ \overline{s_i}   ,~   \underline{s_j}\myPlus \overline{e_j} \myMoins \underline{s_i} ]$& E\\
\scriptsize
$\overline{s_j} \myLe \overline{s_i} \myLe \underline{e_j} \myLe \underline{e_i}$
&& $[ \overline{s_i}   ,~   \underline{e_j}]$ &F\\
  \scriptsize
$\overline{e_j} \myLe \overline{e_i} \myWedge
\overline{e_j} \myGe \overline{s_i} \myWedge
\overline{e_j} \myGe \underline{e_i} \myWedge
\underline{s_i} \myPlus \overline{e_i} \myLe \overline{s_j} \myPlus \underline{e_j}$
&&$[  \underline{s_i} \myPlus \overline{e_i} \myMoins \overline{e_j} , ~  \overline{e_j}]$ &G\\
  \scriptsize
$\underline{e_j} \myLe \overline{e_i}\myWedge
\underline{e_j} \myGe \overline{s_i}\myWedge
\underline{e_j}  \myGe \underline{e_i}\myWedge
\underline{s_i} \myPlus \overline{e_i}  \myLe \underline{s_j} \myPlus \overline{e_j} \myWedge 
\underline{s_i} \myPlus \overline{e_i}  \myLe 2\times \underline{e_j}$
&&$[  \underline{s_i} \myPlus \overline{e_i} \myMoins \underline{e_j} , ~  \underline{e_j}]$ &H\\ \hline
\end{tabular}
\end{minipage}
\vspace{-0.7cm}
\end{table}
}

We have characterized for an activity $a$ the possible value at which condition (\ref{cond:derivativeLT2}) holds. By symmetry, we can deduce the value of the starting time at which condition (\ref{cond:derivativeLT1}) holds.
Table 1 summarizes the intervals of interest of a pair of activities $(i, j)$. 

Our characterization is sharper than the one proposed by Schwindt: For instance, case \emph{B} in Table \ref{tab:characterization} is sharper than the 
equivalent case \emph{iii} in table 3.5 page 84 in \cite{schwindt98:PhD}.
We may notice that intervals of interest may only start at values of the form $\underline{s_a}, \overline{s_a}$ or end at values $\underline{e_a}, \overline{e_a}$.
This leads to a  lighter algorithm for the $O(n^2)$ checker proposed by Baptiste et al. \cite{baptiste:inria-00123562}.
We present the checker for interval with starting dates in $O_S = \{\underline{s_a}, \forall a \in\mathcal{A}\}   \! \cup\!   \{\overline{s_a}, \forall a \in\mathcal{A}\}$  as possible starting values (as CuSP is symmetric).

We define an event as a pair (time,activity). Let $\mathcal{E}_M$ be the events for the last completion times,
$(\overline{e_a},a) \forall a\!\in\!\mathcal{A}$ and ordered in increasing order of time. 
Similarly, we define 
$\mathcal{E}_m = (\underline{e_a},a), \forall a\!\in\!\mathcal{A}$,
$\mathcal{S}_M = (\overline{e_a},a), \forall a\!\in\! \mathcal{A}$ and
$\mathcal{L} = (\underline{s_a}+\overline{e_a},a), \forall a\!\in\!\mathcal{A}$.
\vspace{-0.5cm}
\begin{figure}[h]
\begin{minipage}{0.6\textwidth}  
\begin{algorithm}[H]
{\scriptsize
	\ForEach{$t_1 \in O_S$ }{
		$slope = C-\sum_{a} \MI{t_1}{t_1+1}{a}$\;
		$Load = 0$; ~  $t_2^{old} = t_1$\;
		$\mathcal{L'} = \{ (t' - t_1, a) \mid (t', a) \in {\cal L} \} $\;
		\ForEach{$event (t_2,a) $ in $\mathcal{S}_M$, $\mathcal{E}_m$, $\mathcal{E}_M$, $\mathcal{L'}$}{
			$Load~ +\!\!= slope \times(t_2-t_2^{old})$   \;
			\lIf{$Load<0$}{Fail\;}
			\uIf{event is a $\SOI{a}$}{   \label{algo:ERC:line:ifIsSOI}
				$slope~ -\!\!= h_a$\;
			}			
			\ElseIf{event is a $\DOI{a}$}{  \label{algo:ERC:line:ifIsDOI}
				$slope~ +\!\!= h_a$\;
			}
			$t_2^{old} = t_2$\;
		}
	}
	\caption{Energetic Reasoning Checker.}
	\label{algo:ERChecker}
}
\end{algorithm}
\end{minipage}
\begin{minipage}{0.4\textwidth}  

// The slope represents the evolution of the Slack over time. \\
// Events are evaluated in increasing order of their time. Starting on $t_1\!+\!1$.

\end{minipage}
\vspace{-1cm}
\end{figure}

\section{Experiments and Future Work}
Experiments were run on a 2.9 GHz Intel Core i7, in Choco \cite{choco2010} version 3 (release 13.03).
In order to check the gain obtained with  the new characterization we have considered random instances and instances from the PSPLIB\cite{Kolisch96psplib}. 
Random instances have either 10 or 20 activities. Their processing times were chosen within $[1,10]$, their heights within $[1,5]$.
We used the \emph{first fail} \cite{haralick:ai80} search strategy, and three checkers:  Algorithm~\ref{algo:ERChecker}, Baptiste et al. checker  and the basic $O(n^3)$ checker. 
The number of nodes is identical for all instances  for the three checkers, as expected.
Table~\ref{tab:result:1} shows a running time improvement of  20 to 30\% using Algorithm~\ref{algo:ERChecker}, in comparison with Baptiste et al. algorithm.

We also used our checker in combination of the Time-Table filtering algorithm \cite{Letort:ScalableSweep}, to compare that combination with Time-Table Edge-Finding filtering algorithm~\cite{Vilim2011TTEF}.  
We fixed a solving time limit of five minutes. 
Surprisingly, when proving optimality on random instances with a single resource, using the checker proved optimality for 72 of the 100 generated instances, while TTEF was unable to do so. 
This proves that associating this new checker with a Time-Table approach could be a good default propagator in constraint solvers. 
\vspace{-0.5cm}
\begin{table}[!ht]
{\scriptsize
\begin{minipage}{.5\textwidth}
\centering
\begin{tabular}{|l|r|r|r|}\cline{2-4}
\multicolumn{1}{r|}{}& Algorithm~\ref{algo:ERChecker} & Baptiste et al. &$O(n^3)$ \\
\multicolumn{1}{l|}{Instances} & $(\mu s/node)$ & $(\mu s/node)$ & $(\mu s/node)$ \\ \hline
Random10 &16.47&24.97&29.31\\ 
Random20 &43.95&56.24&78.74\\ \hline
PspLib 30 &450.67&618.77&1268.92\\ 
PspLib 120 &1 339.24&1 683.26&11 288.54\\ \hline
\end{tabular}
\captionof{table}{\label{tab:result:1}Average time per node.}
\end{minipage}
\begin{minipage}{.5\textwidth}
\centering
\begin{tabular}{|l|r|r|r|}\cline{2-4}
\multicolumn{1}{r|}{}&\multicolumn{1}{c|}{TT}  &\multicolumn{1}{c|}{TT + }&\multicolumn{1}{c|}{TT +} \\
\multicolumn{1}{l|}{} &  & TTEF & Algorithm~\ref{algo:ERChecker}  \\ \hline
Random20 &6&7&72\\ \hline
\end{tabular}
\captionof{table}{\label{tab:result:2} \#proved optimum over 100 instances.}
\end{minipage}
}
\end{table}

\vspace{-1.1cm}
As a future work, our results should be confirmed by using dedicated search heuristics, which was not the case in our preliminary experiments. 
Moreover, as a direct consequence of our new case-based characterization, we can answer affirmatively to the second theoretical open question: The interval of interests of Baptiste et al. is complete for the propagation algorithm. As the proof needs a better enlightenment 
and cannot easily fit in a short paper, we will present it in a new paper. 
\linespread{0.96}
\bibliography{report}
\end{document}